\pdfoutput=1

\documentclass[11pt,twoside,a4paper,cmspaper,final,collab]{cms-tdr}
\def\svnVersion{426689P}\def\cmsCernNoTag{CERN-EP-2017-128}\def\cmsCernDate{\today}\def\cmsMessage{Published in the European Physical Journal C as \href{http://dx.doi.org/10.1140/epjc/s10052-017-5192-z}{\doi{10.1140/epjc/s10052-017-5192-z.}}}

\begin{document}\cmsNoteHeader{B2G-17-002}

\hyphenation{had-ron-i-za-tion}
\hyphenation{cal-or-i-me-ter}
\hyphenation{de-vices}
\RCS$Revision: 422991 $
\RCS$HeadURL: svn+ssh://svn.cern.ch/reps/tdr2/papers/B2G-17-002/trunk/B2G-17-002.tex $
\RCS$Id: B2G-17-002.tex 422991 2017-08-30 09:40:07Z zucchett $
\newlength\cmsFigWidth
\ifthenelse{\boolean{cms@external}}{\setlength\cmsFigWidth{0.495\textwidth}}{\setlength\cmsFigWidth{0.7\textwidth}}
\ifthenelse{\boolean{cms@external}}{\providecommand{\cmsLeft}{upper\xspace}}{\providecommand{\cmsLeft}{left\xspace}}
\ifthenelse{\boolean{cms@external}}{\providecommand{\cmsRight}{lower\xspace}}{\providecommand{\cmsRight}{right\xspace}}

\newcommand{\X}{\ensuremath{\cmsSymbolFace{X}}\xspace}
\newcommand{\V}{\ensuremath{\cmsSymbolFace{V}}\xspace}
\renewcommand{\PW}{\ensuremath{\cmsSymbolFace{W}}\xspace}
\newcommand{\htobb}{\ensuremath{\PH\to\bbbar}\xspace}
\newcommand{\VH}{\ensuremath{\V\PH}\xspace}
\newcommand{\VV}{\ensuremath{\V\V}\xspace}
\newcommand{\mVH}{\ensuremath{m_{\VH}}\xspace}
\newcommand{\mX}{\ensuremath{m_{\X}}\xspace}
\newcommand{\mj}{\ensuremath{m_{\mathrm{j}}}\xspace}
\newcommand{\nsub}{\ensuremath{\tau_{21}}\xspace}
\newcommand{\B}{\ensuremath{\mathcal{B}}}
\newcommand{\gV}{\ensuremath{g_\text{V}}\xspace}
\newcommand{\cH}{\ensuremath{c_\text{H}}\xspace}
\newcommand{\cF}{\ensuremath{c_\text{F}}\xspace}
\providecommand{\MGvATNLO}{\MADGRAPH{5}\_a\MCATNLO}
\ifthenelse{\boolean{cms@external}}{\providecommand{\cmsBoldTitle}{{\boldmath $\sqrt{s} = $13}\TeV}}{\providecommand{\cmsBoldTitle}{$\sqrt{s} = 13$\TeV}}

\cmsNoteHeader{B2G-17-002}
\title{Search for heavy resonances that decay into a vector boson and a Higgs boson in hadronic final states at \cmsBoldTitle}
\titlerunning{Search for heavy resonances that decay into a vector boson and a Higgs at 13\TeV}

\date{\today}

\abstract{
A search for heavy resonances with masses above 1\TeV, decaying to final states containing a vector boson and a Higgs boson, is presented. The search considers hadronic decays of the vector boson, and Higgs boson decays to b quarks. The decay products are highly boosted, and each collimated pair of quarks is reconstructed as a single, massive jet. The analysis is performed using a data sample collected in 2016 by the CMS experiment at the LHC in proton-proton collisions at a center-of-mass energy of 13\TeV, corresponding to an integrated luminosity of 35.9\fbinv. The data are consistent with the background expectation and are used to place limits on the parameters of a theoretical model with a heavy vector triplet. In the benchmark scenario with mass-degenerate $\PWpr$ and $\PZpr$ bosons decaying predominantly to pairs of standard model bosons, for the first time heavy resonances for masses as high as 3.3\TeV are excluded at 95\% confidence level, setting the most stringent constraints to date on such states decaying into a vector boson and a Higgs boson.
}

\hypersetup{%
pdfauthor={CMS Collaboration},%
pdftitle={Search for heavy resonances that decay into a vector boson and a Higgs boson in hadronic final states at sqrt(s)=13 TeV},%
pdfsubject={CMS},%
pdfkeywords={CMS, physics, B2G, diboson, VH, Higgs, hadronic}
}

\maketitle

\section{Introduction}

The discovery of the Higgs boson (\PH) at the CERN LHC~\cite{bib:Aad20121,bib:Chatrchyan201230,Chatrchyan:2013lba} represents a milestone in the understanding of the standard model (SM) of particle physics. However, the degree of fine-tuning required to accommodate the observed mass of 125\GeV~\cite{Aad:2014aba,Khachatryan:2014jba,Chatrchyan:2014vua,Aad:2015zhl} suggests the presence above 1\TeV of new heavy particles beyond the SM (BSM), possibly lying within reach of the LHC.
These resonances, denoted as \X, are expected to be connected to the electroweak sector of the SM, with significant couplings to the SM bosons.
Hence, these heavy resonances potentially could be observed through their decay into a vector boson ($\V = \PW$ or $\Z$) and a Higgs boson.

The VH resonances are predicted in several BSM theoretical models, most notably weakly coupled spin-1 $\PZpr$~\cite{Barger:1980ix,1126-6708-2009-11-068} and $\PWpr$ models~\cite{Grojean:2011vu}, strongly coupled composite Higgs models~\cite{Contino2011,Marzocca2012,Bellazzini:2014yua}, and little Higgs models~\cite{Han:2003wu,Schmaltz,Perelstein2007247}.
The heavy vector triplet (HVT) framework~\cite{Pappadopulo2014} extends the SM by introducing a triplet of heavy vector bosons, one neutral $\PZpr$ and two charged $\PWpr$s, collectively represented as \V' and degenerate in mass. The heavy vector bosons couple to SM bosons and fermions with strengths $\gV\cH$ and $g^2 \cF / \gV$, respectively, where $\gV$ is the strength of the new interaction, $\cH$ is the coupling between the HVT bosons, the Higgs boson, and longitudinally polarized SM vector bosons, $\cF$ is the coupling between the HVT bosons and the SM fermions, and $g$ is the $SU(2)_L$ gauge coupling.
In this paper, two different benchmark scenarios are considered~\cite{Pappadopulo2014}. In model~A ($\gV=1$, $\cH=-0.556$, $\cF=-1.316$), the coupling strengths to the SM bosons and fermions are comparable, and the new particles decay primarily to fermions. In model~B ($\gV = 3$, $\cH=-0.976$, $\cF=1.024$), the couplings to fermions are suppressed with respect to the couplings to bosons, resulting in a branching fraction to SM bosons close to unity.

This paper describes the search in proton-proton collisions at 13\TeV for heavy resonances decaying to final states containing a SM vector boson and a Higgs boson, which subsequently decay into a pair of quarks and a pair of b quarks, respectively. Use of the hadronic decay modes takes advantage of the large branching fractions, which compensate for the effect of the large multijet background.
This search concentrates on the high mass region, as previous searches~\cite{Khachatryan:2015lba,Aad:2015yza,Khachatryan:2015bma,Khachatryan:2015ywa,Sirunyan:2016cao,Aaboud:2016okv,Khachatryan2017137,Aaboud:2016lwx} have excluded \mX in the region below a few \TeV.
As a result of the large resonance mass, the two bosons produced in the decay have large Lorentz boosts in the laboratory frame, and consequently the hadronic decay products of each boson tend to be clustered within a single hadronic jet. The jet mass, substructure, and b tagging information are crucial to identifying hadronically decaying vector bosons and Higgs boson candidates, and to discriminating against the dominant SM backgrounds.

This search complements and significantly extends the reach of the CMS search with 2015 data for VH resonances with semileptonic decay modes of the vector bosons~\cite{Khachatryan2017137}, which excludes at 95\% confidence level (CL) \PWpr and \PZpr resonances with mass below 1.6\TeV and mass-degenerate \V' resonances with masses up to 2.0\TeV in the HVT benchmark model B. The ATLAS Collaboration has performed a search in the same final state with a comparable data set, excluding \PWpr and \PZpr bosons with masses below 2.2 and 1.6\TeV, respectively, and a \V' boson with mass below 2.3\TeV in the HVT model~B scenario~\cite{Aaboud:2016lwx}.

\section{Data and simulated samples}\label{sec:mcsimulation}

The data sample studied in this analysis was collected in 2016 with the CMS detector in proton-proton collisions at a center-of-mass energy of 13\TeV, and corresponds to an integrated luminosity of 35.9\fbinv.

Simulated signal events are generated at leading order (LO) with the \MGvATNLO 2.2.2 matrix element generator~\cite{MadGraph}. The Higgs boson is required to decay into a \bbbar pair, and the vector boson to decay hadronically. Other decay modes are not considered in the present analysis. Different hypotheses for the heavy resonance mass \mX in the range 1000 to 4500\GeV are considered, assuming a narrow resonance width (0.1\% of the mass), which is small with respect to the experimental resolution. This narrow-width assumption is valid in a large fraction of the HVT parameter space, and fulfilled in both benchmark models~A and B~\cite{Pappadopulo2014}.

Although the background is estimated using a method based on data, simulated background samples are generated for the optimization of the analysis selections.
Multijet background events are generated at LO with \MGvATNLO, and top quark pair production is simulated at next-to-leading order (NLO) with the \POWHEG 2.0 generator~\cite{Nason:2004rx,Frixione:2007vw,Alioli:2010xd} and rescaled to the cross section computed with \textsc{Top++} v2.0~\cite{Czakon:2011xx} at next-to-next-to-leading order.
Other SM backgrounds, such as W+jets, Z+jets, single top quark production, \VV, and nonresonant \VH production, are simulated at NLO in QCD with \MGvATNLO using the {FxFx} merging scheme~\cite{Frederix:2012ps}.
Parton showering and hadronization processes are interfaced with~\PYTHIA 8.205~\cite{Sjostrand:2007gs} with the CUETP8M1 underlying event tune~\cite{Skands:2014pea,Khachatryan:2015pea}. The CUETP8M2T4 tune~\cite{CMS-PAS-TOP-16-021} is used for top quark pair production.
The NNPDF 3.0~\cite{Ball:2014uwa} parton distribution functions (PDFs) are used in generating all simulated samples.
Additional collisions in the same or adjacent bunch crossings (pileup) are taken into account by superimposing simulated minimum bias interactions onto the hard scattering process, with a frequency distribution matching that observed experimentally. The generated events are processed through a full detector simulation based on {\GEANTfour}~\cite{Agostinelli:2002hh} and reconstructed with the same algorithms as used for collision data.

\section{The CMS detector}\label{sec:detector}

The central feature of the CMS detector is a superconducting solenoid with a 6\unit{m} internal diameter. In the solenoid volume, a silicon pixel and strip tracker measures charged particles within the pseudorapidity range $\abs{\eta}< 2.5$. The tracker consists of 1440 silicon pixel and 15\,148 silicon strip detector modules and is located in the 3.8\unit{T} field of the solenoid. For nonisolated particles of transverse momentum $1 < \pt < 10\GeV$ and $\abs{\eta} < 1.4$, the track resolutions are typically 1.5\% in \pt and 25--90 (45--150)\mum in the transverse (longitudinal) impact parameter \cite{Chatrchyan:2014fea}. A lead tungstate crystal electromagnetic calorimeter (ECAL), and a brass and scintillator hadron calorimeter (HCAL), each composed of a barrel and two endcap sections, provide coverage up to $\abs{\eta} < 3.0$, which is further extended by forward calorimeters. Muons are measured in drift tubes, cathode strip chambers, and resistive-plate chambers embedded in the steel flux-return yoke outside the solenoid.

The first level of the CMS trigger system~\cite{Khachatryan:2016bia}, composed of custom hardware processors, uses information from the calorimeters and muon detectors to select the most interesting events in a fixed time interval of less than 4\mus. The high-level trigger (HLT) processor farm decreases the event rate from around 100\unit{kHz} to about 1\unit{kHz}, before data storage.

A detailed description of the CMS detector, together with a definition of the coordinate system used and the relevant kinematic variables, can be found in Ref.~\cite{Chatrchyan:2008zzk}.

\section{Event reconstruction}\label{sec:eventreconstruction}

The event reconstruction employs a particle-flow (PF) algorithm~\cite{CMS-PAS-PFT-09-001,CMS-PAS-PFT-10-001}, which uses an optimized combination of information from the various elements of the CMS detector to reconstruct and identify individual particles produced in each collision. The algorithm identifies each reconstructed particle either as an electron, a muon, a photon, a charged hadron, or a neutral hadron.
The PF candidates are clustered into jets using the anti-\kt algorithm~\cite{Cacciari:2008gp,Cacciari:2011ma} with a distance parameter $R=0.8$, after passing the charged-hadron subtraction (CHS) pileup mitigation algorithm~\cite{CMS-PAS-JME-14-001}.
For each event, a primary vertex is identified as the one with the highest sum of the $\pt^2$ of the associated reconstructed objects, jets and identified leptons, and missing transverse momentum. The CHS algorithm removes charged PF candidates with a track longitudinal impact parameter not compatible with this primary vertex.
The contribution to a jet of neutral particles originating from pileup interactions, assumed to be proportional to the jet area~\cite{Cacciari:2008gn}, is subtracted from the jet energy.
Jet energy corrections as a function of the \pt and $\eta$ are extracted from simulation and data in dijet, multijet, $\gamma$+jets, and leptonic Z+jets events.
The jet energy resolution typically amounts to 5\% at 1\TeV~\cite{Chatrchyan:2011ds,Khachatryan:2016kdb}.
Jets are required to pass identification criteria in order to remove spurious jets arising from detector noise~\cite{CMS:2016ljj}. This requirement has negligible impact on the signal efficiency.

Although AK8 CHS jets are considered for their kinematic properties, the mass of the jet and the substructure variables are determined with a more sophisticated algorithm than the CHS procedure, denoted as pileup-per-particle identification (PUPPI) \cite{Bertolini2014}. The PUPPI algorithm uses a combination of the three-momenta of the particles, event pileup properties, and tracking information in order to compute a weight, assigned to charged and neutral candidates, describing the likelihood that each particle originates from a pileup interaction.
The weight is used to rescale the particle four-momenta, superseding the need for further jet-based corrections.
The PUPPI constituents are subsequently clustered with the same algorithm used for CHS jets, and then are matched with near 100\% efficiency to the AK8 jets clustered with the CHS constituents.

The soft-drop algorithm~\cite{Dasgupta:2013ihk,Larkoski:2014wba}, which is designed to remove contributions from soft radiation and additional interactions, is applied to PUPPI jets. The angular exponent parameter of the algorithm is set to $\beta = 0$, and the soft threshold to $z_\text{cut} = 0.1$. The soft-drop jet mass is defined as the invariant mass associated with the four-momentum of the jet after the application of the soft-drop algorithm.
Dedicated mass corrections, derived from simulation and data in a region enriched with $\ttbar$ events having merged $\PW(\qqbar)$ decays, are applied to each jet mass in order to remove any residual jet \pt dependence~\cite{CMS-PAS-JME-16-003}, and to match the jet mass scale and resolution observed in data. The measured jet mass resolution, obtained after applying the PUPPI and soft-drop algorithms, is approximately 10\%.

Substructure variables are used to identify single reconstructed jets that result from the merger of more than one parton jet. These variables are calculated on each reconstructed jet before the application of the soft-drop algorithm including the PUPPI algorithm corrections for pileup mitigation. The constituents of the jet are clustered iteratively with the anti-\kt algorithm, and the procedure is stopped when $N$ subjets are obtained. A variable, the $N$-subjettiness~\cite{Thaler:2010tr}, is introduced:
$$\tau_N = \frac{1}{d_0} \sum_k p_{\mathrm{T},k} \, \min( \Delta R_{1,k}, \Delta R_{2,k}, \dots, \Delta R_{N,k} ).$$
The index $k$ runs over the jet constituents and the distances $\Delta R_{J,k}$ are calculated with respect to the axis of the $J$th subjet.
The normalization factor $d_0$ is calculated as $d_0 = \sum_k p_{\mathrm{T}, k} R_0$, setting $R_0$ to the radius of the original jet.
The variable that best discriminates between quark and gluon jets and jets from two-body decays of massive particles is the ratio of 2-subjettiness and 1-subjettiness, $\nsub = \tau_2 / \tau_1$, which lies in the interval from 0 to 1, where small values correspond to a high compatibility with the hypothesis of a massive object decaying into two quarks.
The normalization scale factors relative to the \nsub categories are measured from data in a sample enriched in \ttbar events in two \nsub intervals ($0.99 \pm 0.11$ for $\nsub < 0.35$, and $1.03 \pm 0.23$ for $0.35 < \nsub < 0.75$)~\cite{CMS-PAS-JME-16-003}. These two selections are approximately 50 and 45\% efficient for identifying two-pronged jets produced in a decay of a massive boson, and 10 and 60\% efficient on one-pronged jets, respectively. The threshold values are chosen in order to maximize the overall sensitivity over the entire mass spectrum.

The Higgs boson jet candidates are identified using a dedicated b tagging discriminator, specifically designed to identify a pair of b quarks clustered in a single jet~\cite{CMS-PAS-BTV-15-002}. The algorithm combines information from displaced tracks and the presence of one or two secondary vertices within the Higgs boson jet in a dedicated multivariate algorithm. The decay chains of the two b hadrons are resolved by associating reconstructed secondary vertices with the directions of the two $N$-subjettiness axes.
Tight and loose operating points are chosen for Higgs boson jets that have corresponding false-positive rates for light quark and gluon jets being identified as jets from b quarks of about 0.8 and 8\%, with efficiencies of approximately 35 and 75\%, respectively.
Scale factors, derived from data in events enriched by jets containing muons~\cite{CMS-PAS-BTV-15-002}, are applied to the simulation to correct for the differences between data and simulation.

Since the analysis concentrates on hadronic final states, events containing isolated charged leptons or large missing transverse momentum are rejected.
Electrons are reconstructed in the fiducial region $\abs{\eta}<2.5$ by matching the energy deposits in the ECAL with tracks reconstructed in the tracker~\cite{Khachatryan:2015hwa}.
Muons are reconstructed within the acceptance of the CMS muon systems, $\abs{\eta}<2.4$, using the information from both the muon spectrometer and the silicon tracker~\cite{Chatrchyan:2012xi}.
The isolation of electrons and muons is based on the summed energy of reconstructed PF candidates within a cone around the lepton direction.
Hadronically decaying $\tau$ leptons are reconstructed in the $\abs{\eta}<2.3$ region by combining one or three hadronic charged PF candidates with up to two neutral pions, the latter also reconstructed by the PF algorithm from the photons arising from the $\Pgpz \to\gamma\gamma$ decay~\cite{Khachatryan:2015dfa}.
The missing transverse momentum is calculated as the magnitude of the vector sum of the momenta of all PF candidates projected onto the plane perpendicular to the beams.

\section{Event selection}
\label{sec:evtsel}

Events are collected with four triggers~\cite{Khachatryan:2016bia}. The first requires \HT, defined as the scalar sum of the transverse momentum of the PF jets, to be larger than 800 or 900 \GeV, depending on the instantaneous luminosity. The second trigger, with a lower \HT threshold set to 650 \GeV, is also required to have a pair of PF jets with invariant mass larger than 950\GeV, and pseudorapidity separation $\abs{\Delta \eta}$ smaller than 1.5. A third trigger requires at least one PF jet with \pt larger than 450 \GeV. The fourth trigger selects events with at least one PF jet with $\pt> 360\GeV$ passing a trimmed mass~\cite{Krohn:2009th} threshold of 30\GeV, or $\HT>700\GeV$ and trimmed mass larger than 50 \GeV. In all these triggers, reconstruction of PF jets is based on the anti-\kt algorithm with $R=0.4$, rather than $R=0.8$ as used offline.

\begin{figure}[!hbtp]\centering
  \includegraphics[width=0.495\textwidth]{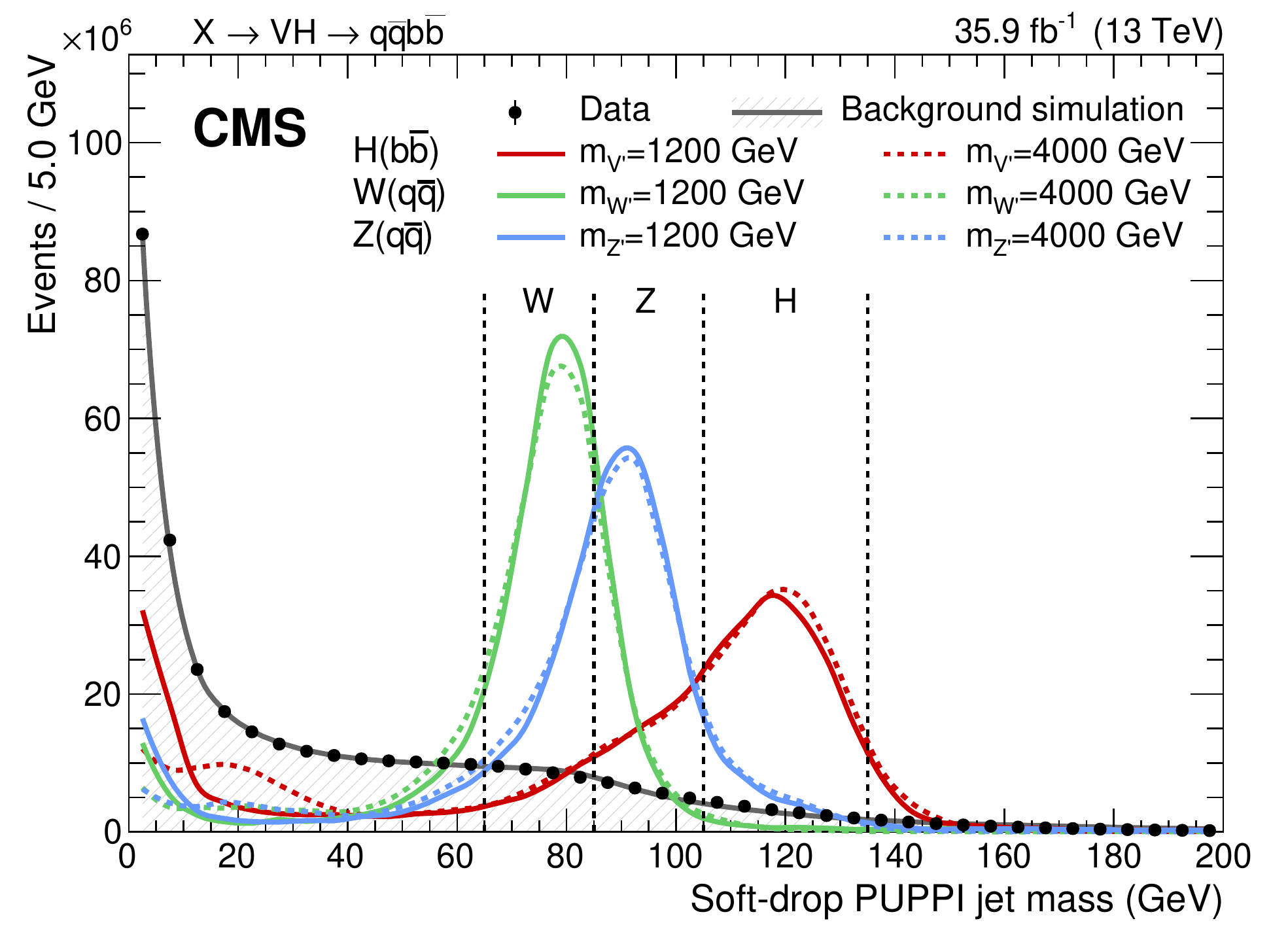}
  \caption{
    Distribution of the soft-drop PUPPI mass after the kinematic selections on the two jets, for data, simulated background, and signal. The signal events with low mass correspond to boson decays where one of the two quarks is emitted outside the jet cone or the two quarks are overlapping. The distributions are normalized to the number of events observed in data. The dashed vertical lines represent the boundaries between the jet mass categories.
    \label{fig:var1} }
\end{figure}

In the offline preselection, the two jets with highest \pt in the event are required to have $\pt>200\GeV$ and $\abs{\eta}<2.5$, and $\abs{\Delta \eta} \leq 1.3$. At least one of the two jets must have a soft-drop jet mass compatible with the Higgs boson mass, $105 < \mj < 135\GeV$ (\PH jet), and the other jet a mass compatible with the mass of the vector bosons, $65 < \mj < 105\GeV$ (\V jet). The jet mass categorization is shown in Fig.~\ref{fig:var1}.
The \PH jet and \V jet candidates are required to have a combined invariant mass \mVH larger than 985\GeV, to avoid trigger threshold effects and thus ensure full efficiency.
Events with isolated electrons or muons with $\pt > 10\GeV$, or $\tau$ leptons with $\pt > 18 \GeV$, are rejected. The reconstructed missing transverse momentum is required to be smaller than 250\GeV.

The events passing the preselection are divided into eight exclusive categories. Two categories are defined for the \PH jet, depending on the value of the b tagging discriminator: a \textit{tight} category containing events with a discriminator larger than 0.9, and a \textit{loose} category requiring a value between 0.3 and 0.9. Similarly, two categories of \V jets are defined using the subjettiness ratio: a \textit{high purity} category containing events with $\nsub \leq 0.35$, and a \textit{low purity} category having $0.35 < \nsub < 0.75$.
Although it is expected that the tight and high purity categories dominate the total sensitivity, the loose and low purity categories are retained since for large dijet invariant mass they provide a nonnegligible signal efficiency with an acceptable level of background contamination.

\begin{figure}[!hbtp]\centering
  \includegraphics[width=0.495\textwidth]{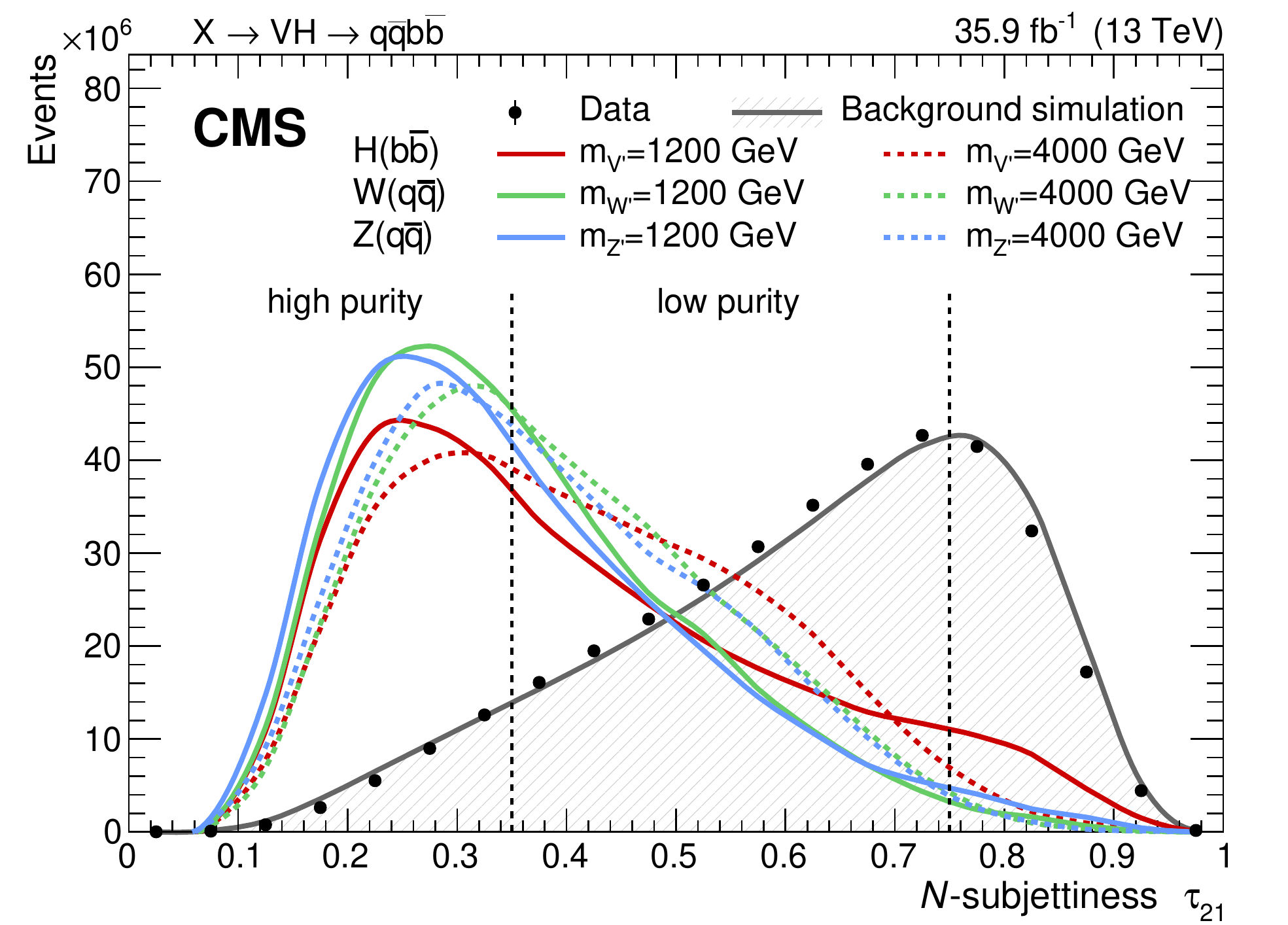}
  \includegraphics[width=0.495\textwidth]{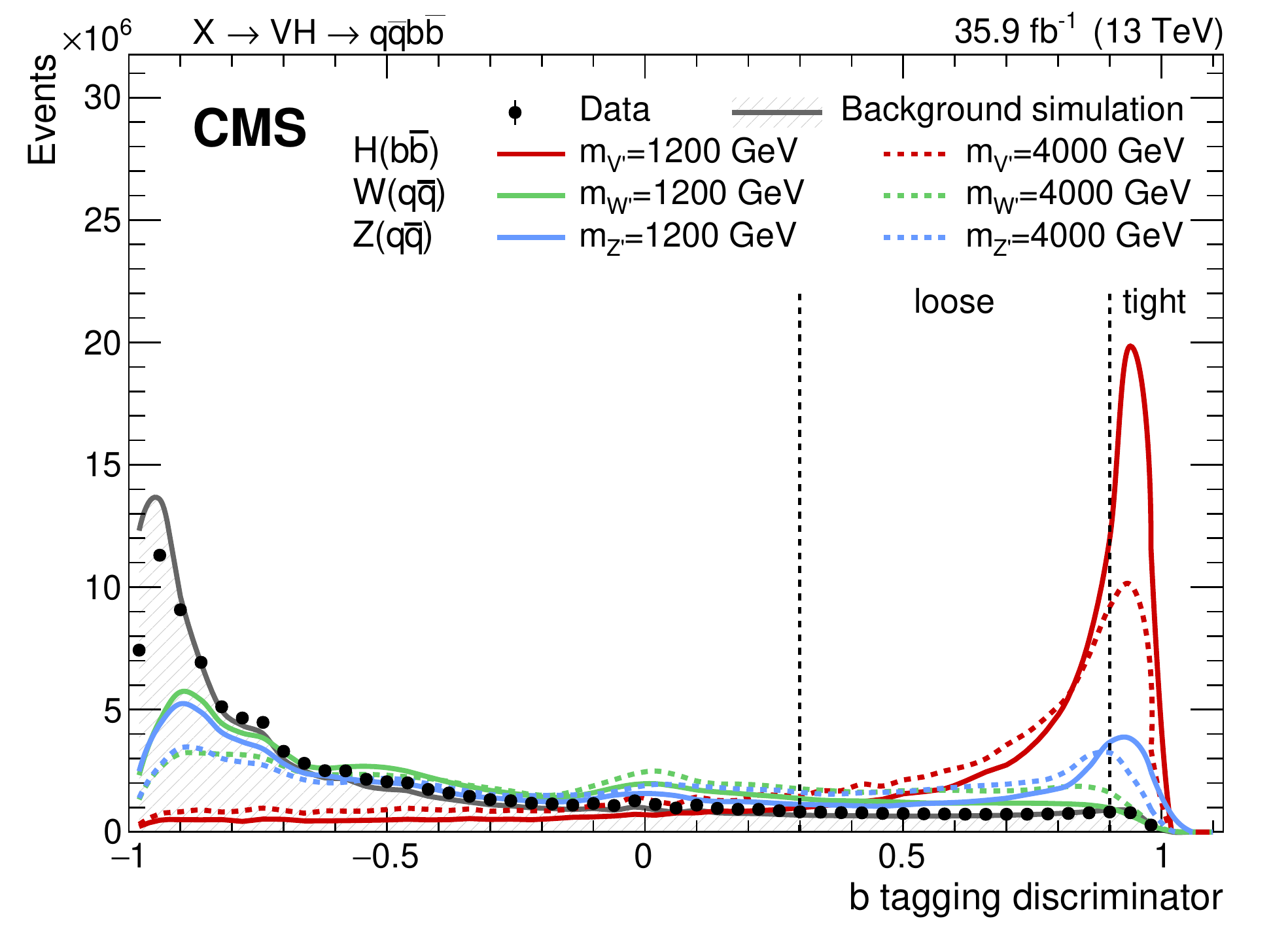}
  \caption{
    Distribution of the $N$-subjettiness \nsub (\cmsLeft) and b tagging discriminator output (\cmsRight) after the kinematic selections on the two jets, for data, simulated background, and signal. The distributions are normalized to the number of events observed in data. The dashed vertical lines represent the boundaries between the categories as described in the text.
    \label{fig:var2} }
\end{figure}

Two further categories are defined based on the \V jet mass, by splitting the mass interval. Events with \V jet mass closer to the nominal \PW boson mass value, $65 < \mj \leq 85\GeV$, are assigned to a W mass category, and those with $85 < \mj \leq 105\GeV$ fall into a Z mass category. Even if the \PW and \Z mass peaks cannot be fully resolved, this classification allows a partial discrimination between a potential \PWpr or \PZpr signal.
The signal efficiency for the combination of the eight categories reaches 36\% at $\mX=1.2$--$1.6\TeV$, and slowly decreases to $21\%$ at $\mX=4.5\TeV$.
The $N$-subjettiness and b tagging categorizations are shown in Fig.~\ref{fig:var2}.

\section{Background estimation}
\label{sec:bkg}

The background is largely dominated by multijet production, which accounts for more than 95\% of the total background. The top quark pair contribution is approximately 3--4\%, depending on the category. The remaining fraction is composed of vector boson production in association with partons, and SM diboson processes.

\begin{figure*}[!hbtp]\centering
  \includegraphics[width=0.495\textwidth]{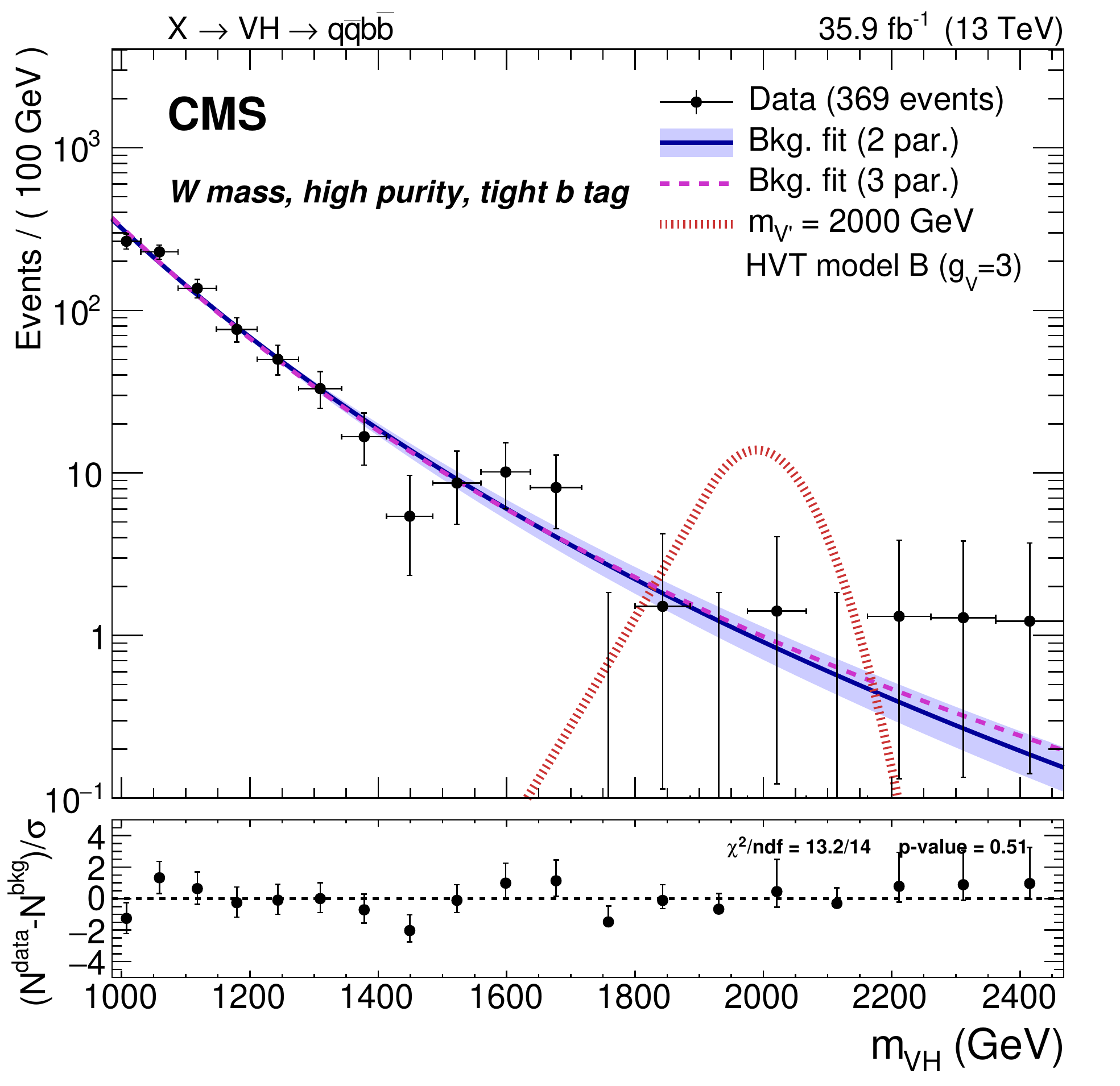}
  \includegraphics[width=0.495\textwidth]{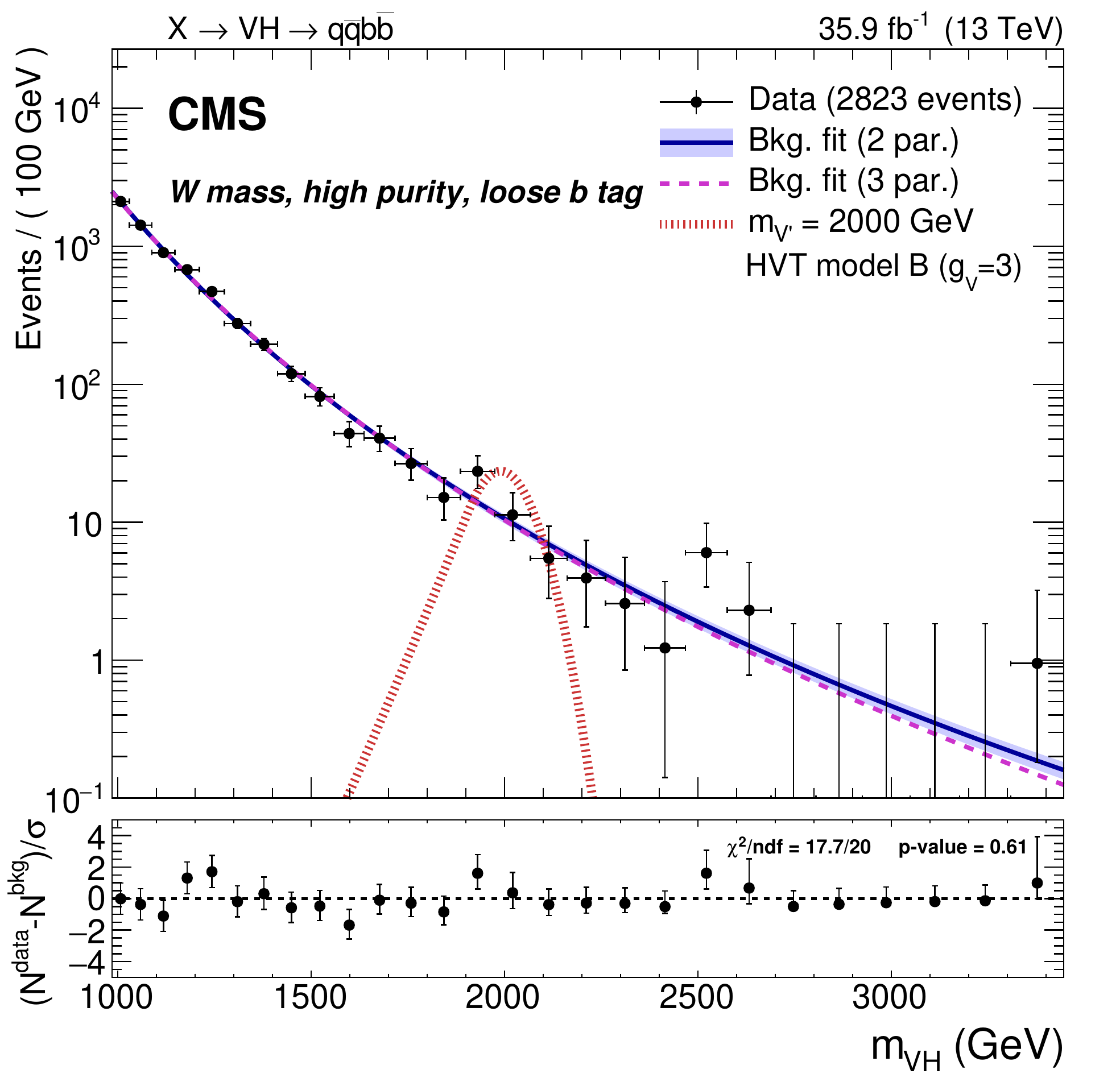}

  \includegraphics[width=0.495\textwidth]{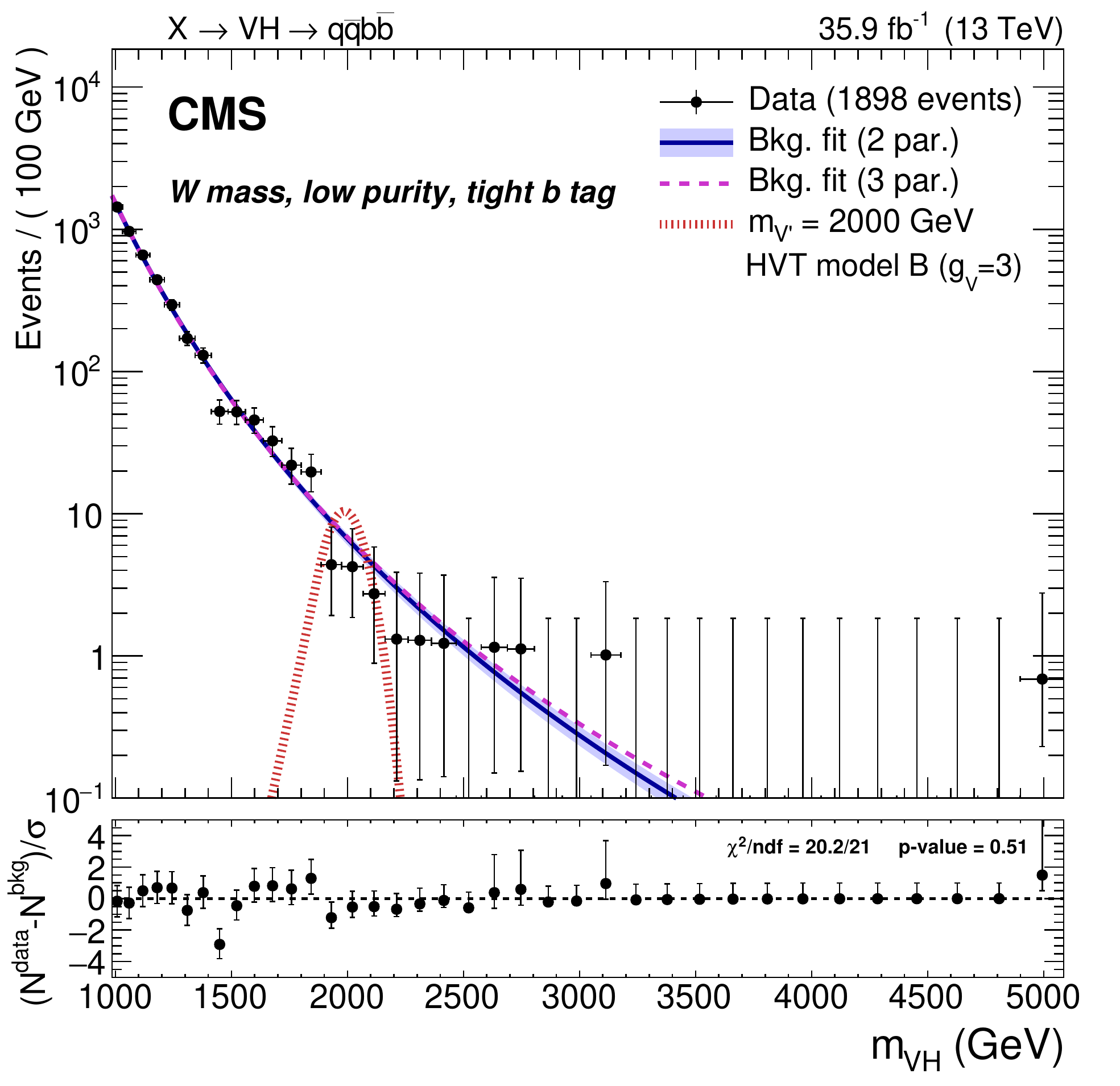}
  \includegraphics[width=0.495\textwidth]{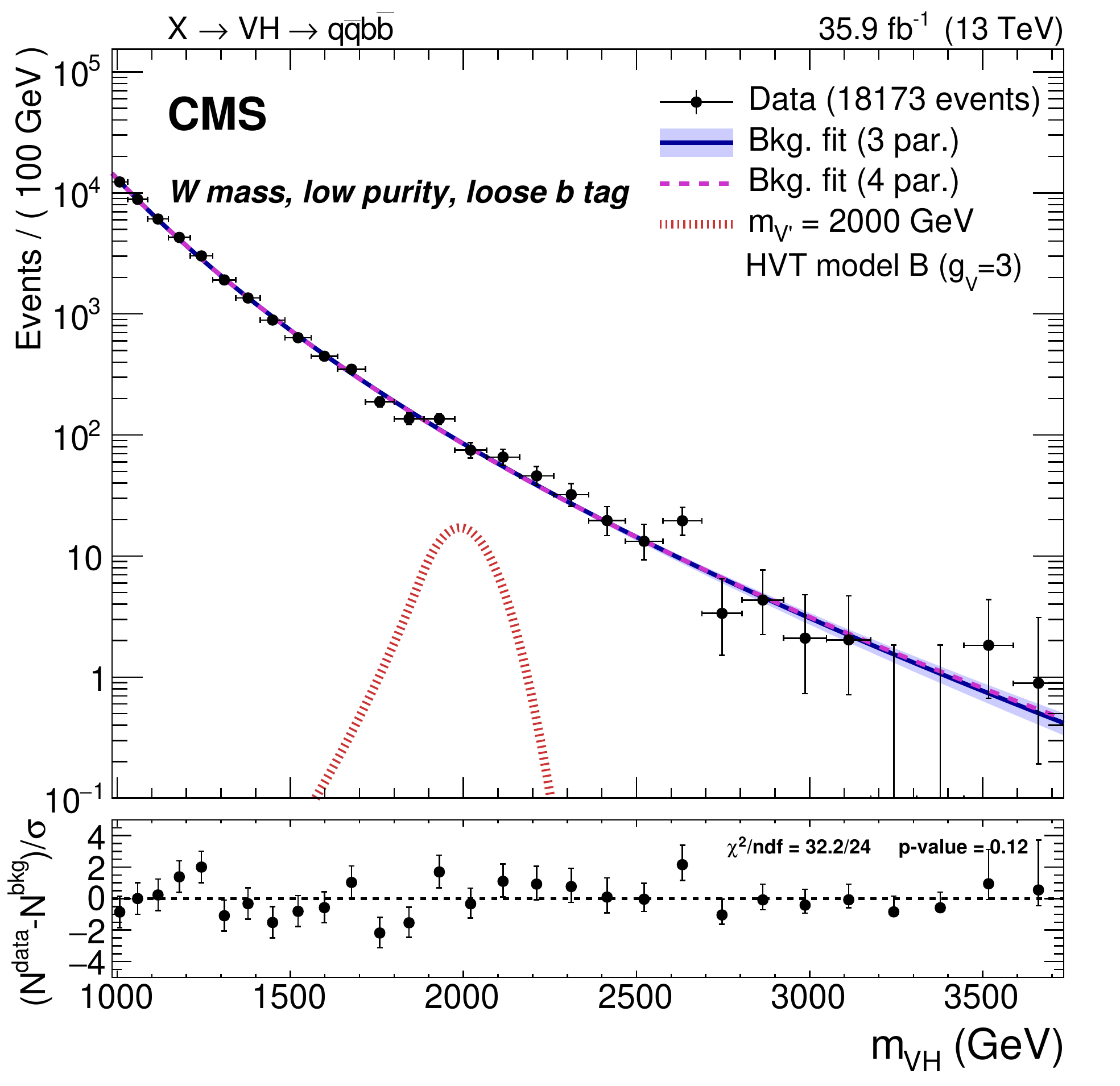}
  \caption{
    Dijet invariant distribution \mVH of the two leading jets in the \PW mass region: high purity (upper) and low purity (lower) categories, with tight (\cmsLeft) and loose (\cmsRight) b tagging selections. The preferred background-only fit is shown as a solid blue line with an associated shaded band indicating the uncertainty. An alternative fit is shown as a purple dashed line. The ratio panels show the pulls in each bin, $(N^\text{data}-N^\text{bkg})/\sigma$, where $\sigma$ is the Poisson uncertainty in data. The horizontal bars on the data points indicate the bin width and the vertical bars represent the normalized Poisson errors, and are shown also for bins with zero entries up to the highest \mVH event. The expected contribution of a resonance with $\mX = 2000\GeV$, simulated in the context of the HVT model B, is shown as a dot-dashed red line.
    \label{fig:wr} }
\end{figure*}

\begin{figure*}[!hbtp]\centering
  \includegraphics[width=0.495\textwidth]{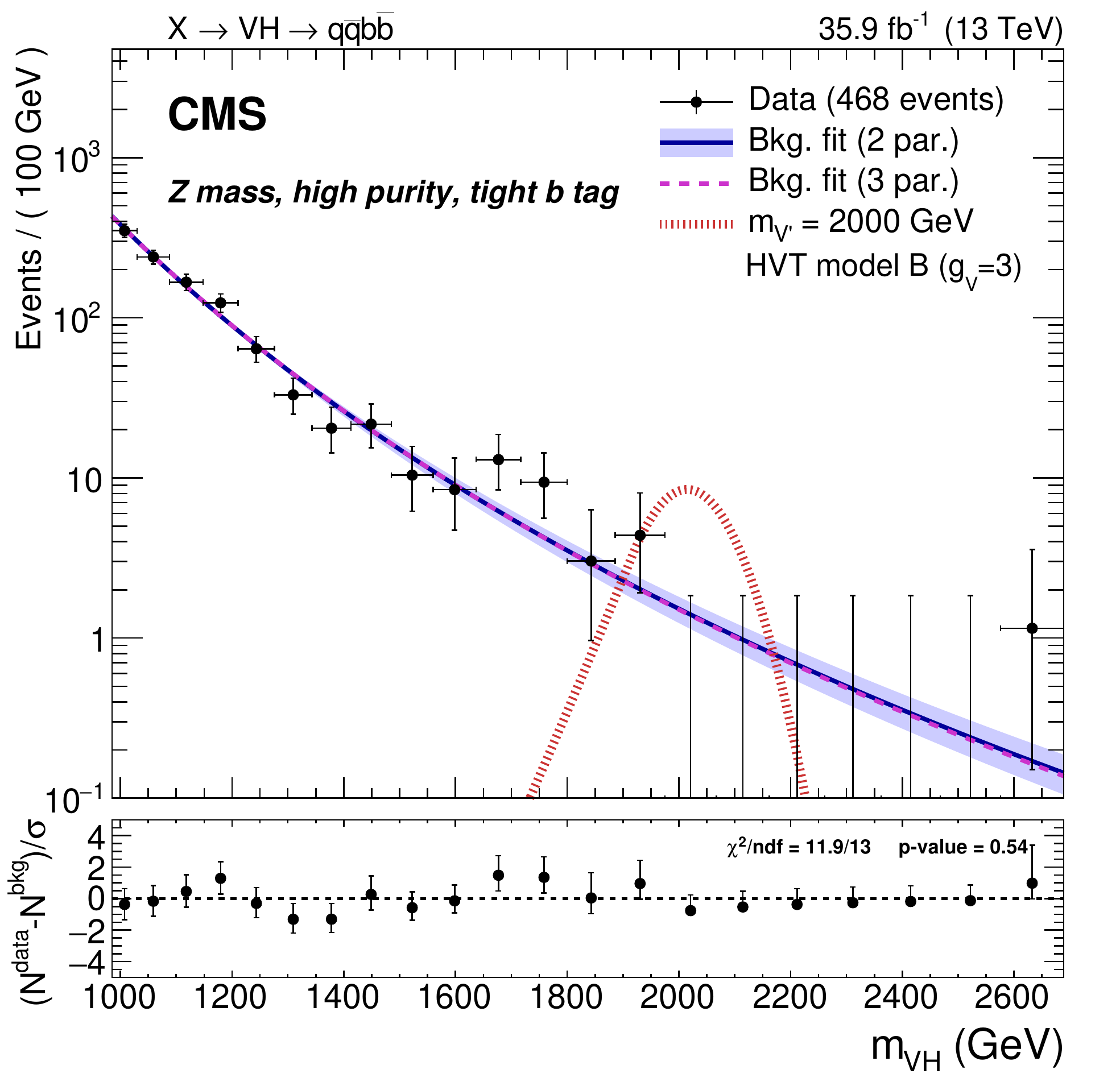}
  \includegraphics[width=0.495\textwidth]{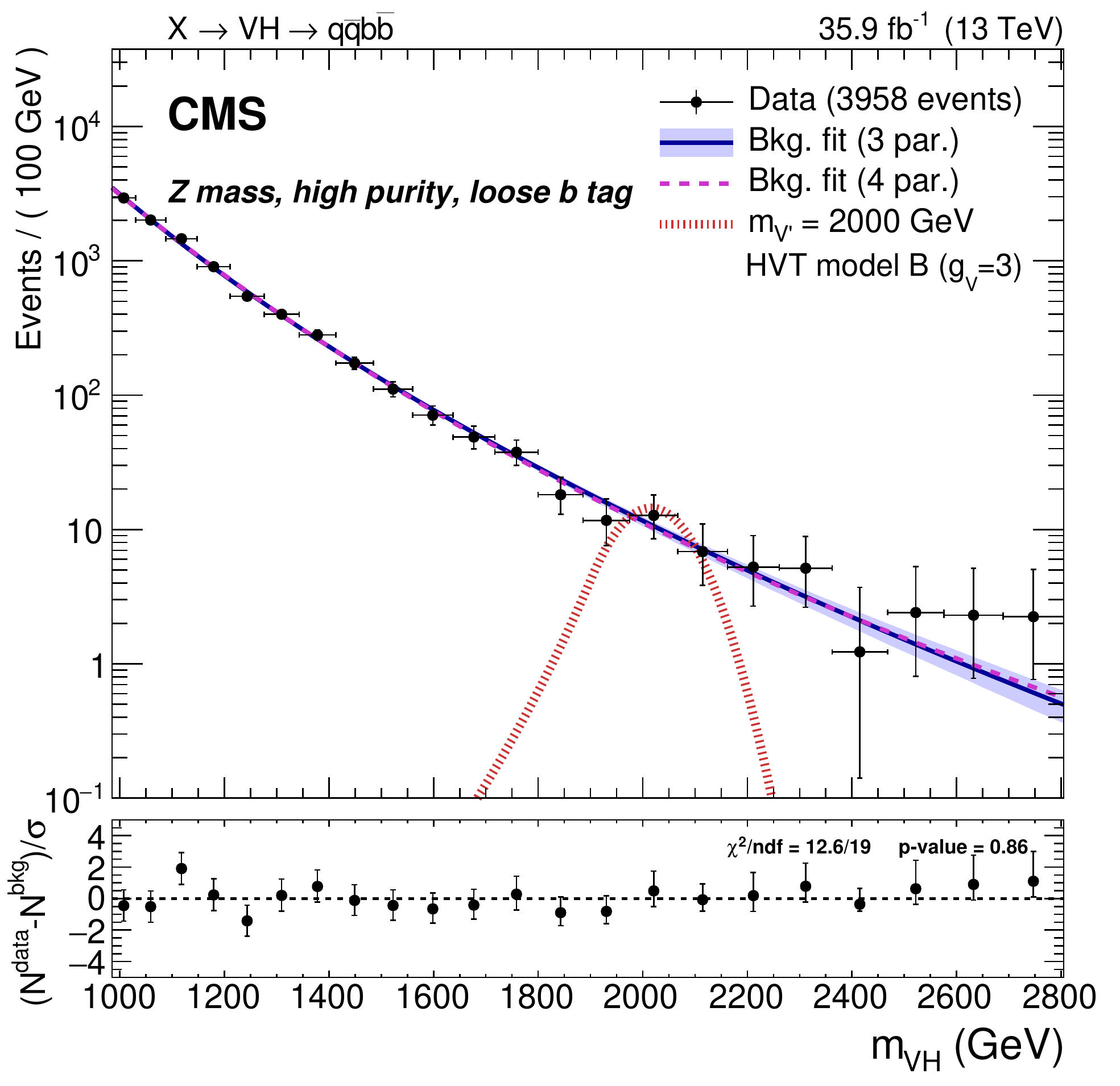}

  \includegraphics[width=0.495\textwidth]{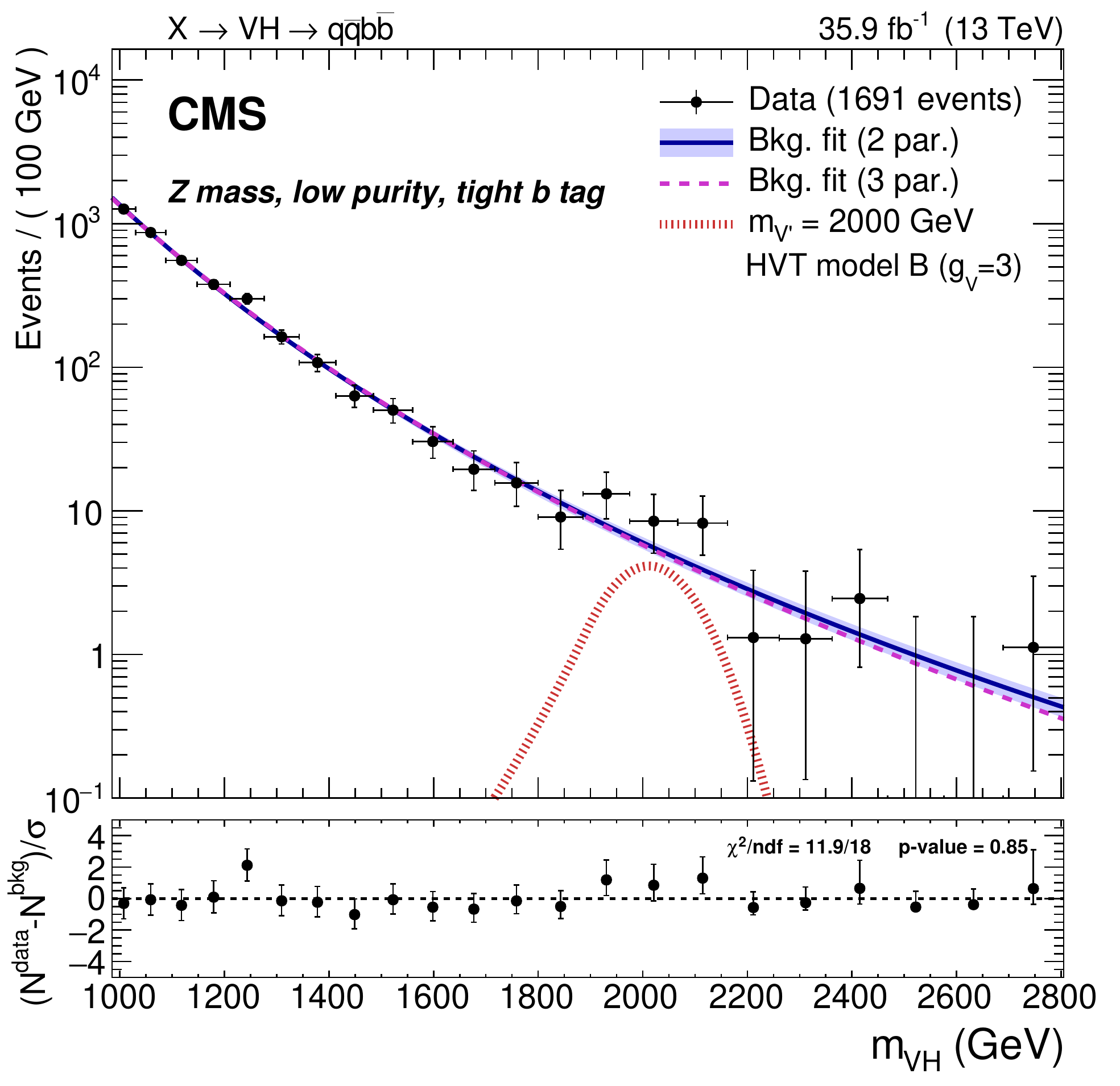}
  \includegraphics[width=0.495\textwidth]{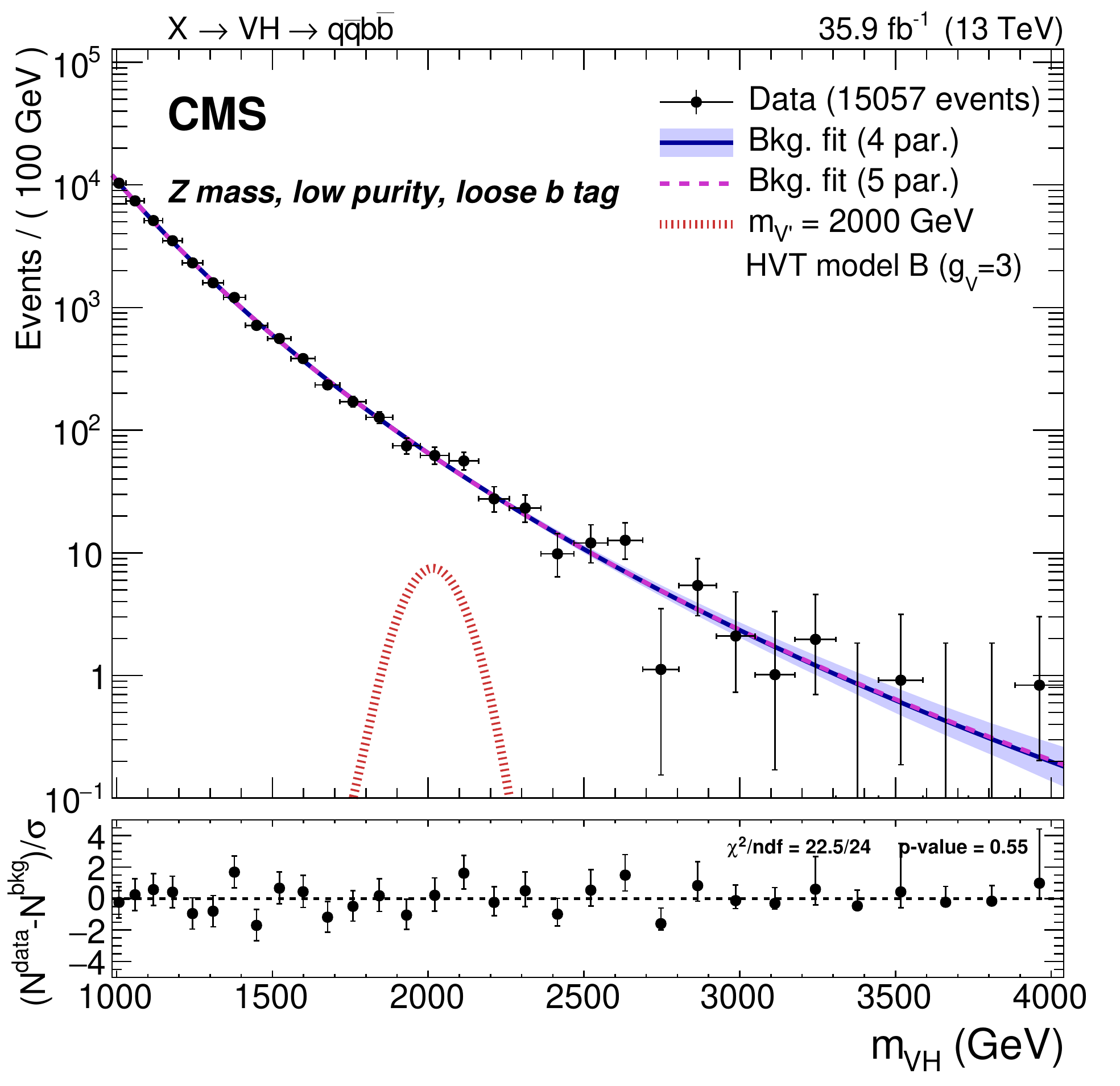}
  \caption{
    Dijet invariant distribution \mVH of the two leading jets in the \Z mass region: high purity (upper) and low purity (lower) categories, with tight (\cmsLeft) and loose (\cmsRight) b tagging selections. The preferred background-only fit is shown as a solid blue line with an associated shaded band indicating the uncertainty. An alternative fit is shown as a purple dashed line. The ratio panels show the pulls in each bin, $(N^\text{data}-N^\text{bkg})/\sigma$, where $\sigma$ is the Poisson uncertainty in data. The horizontal bars on the data points indicate the bin width and the vertical bars represent the normalized Poisson errors, and are shown also for bins with zero entries up to the highest \mVH event. The expected contribution of a resonance with $\mX = 2000\GeV$, simulated in the context of the HVT model B, is shown as a dot-dashed red line.
    \label{fig:zr} }
\end{figure*}

The background is estimated directly from data, assuming that the \mVH distribution can be described by a smooth, parametrizable, monotonically decreasing function. This assumption is verified in the \V jet mass sidebands ($40 < \mj < 65\GeV$) and in simulation.
The expressions considered are functions of the variable $x = \mVH/\sqrt{s}$, where $\sqrt{s} = 13 \TeV$ is the center of mass energy, and the number of parameters $p_i$, including the normalization, is between two and five:
\ifthenelse{\boolean{cms@external}}{
\begin{multline*}
\frac{p_0}{x^{p_1}}, \quad \frac{p_0 \, \left( 1-x \right)^{p_1}}{x^{p_2}}, \quad \frac{p_0 \, \left( 1-x \right)^{p_1}}{x^{p_2 + p_3 \, \log\left( x \right)}}, \\ \frac{p_0 \, \left( 1-x \right)^{p_1}}{x^{p_2 + p_3  \log\left( x \right) + p_4 \, \log^2\left( x \right)}}.
\end{multline*}
}{
\begin{equation*}
\frac{p_0}{x^{p_1}}, \quad \frac{p_0 \, \left( 1-x \right)^{p_1}}{x^{p_2}}, \quad \frac{p_0 \, \left( 1-x \right)^{p_1}}{x^{p_2 + p_3 \, \log\left( x \right)}}, \quad \frac{p_0 \, \left( 1-x \right)^{p_1}}{x^{p_2 + p_3 \, \log\left( x \right) + p_4 \, \log^2\left( x \right)}}.
\end{equation*}
}
Starting from the simplest functional form, an iterative procedure based on the Fisher F-test~\cite{Fisher} is used to check at 10\% CL if additional parameters are needed to model the background distribution.
For most categories, the two-parameter functional form is found to describe the data spectrum sufficiently well. However, in more populated categories, with loose b tagging or low purity, three- or four-parameter functions are preferred.
The results of the fits are shown in Figs.~\ref{fig:wr} and \ref{fig:zr} for the \PW and \Z mass regions, respectively.
Although the fits are unbinned, the binning chosen to present the results is consistent with the detector resolution.
The event with the highest invariant mass observed has $\mVH = 4920\GeV$ and is in the \PW mass, low purity, tight b tag category.

The shape of the reconstructed signal mass distribution is extracted from the simulated signal samples. The signal shape is parametrized separately for each channel with a Gaussian peak and a power law to model the lower tail, for a total of four parameters. The reconstruction resolution for \mVH is taken to be the width of the Gaussian core, and is 4\% at low resonance mass and 3\% at high mass.

Dedicated tests have been performed to check the robustness of the fit method by generating pseudo-experiments after injecting a simulated signal with various mass values and cross sections on top of the nominal fitted function. The pseudo-data distribution is then subjected to the same procedure as the data, including the F-test, to determine the background function.
The signal yield derived from a combined background and signal fit is found to be compatible with the injected yield within one third of the statistical uncertainty, regardless of the injected signal strength and resonance mass. These tests verify that the possible presence of a signal and the choice of the function used to model the background do not introduce significant biases in the final result.

\section{Systematic uncertainties}
\label{sec:sys}

The background estimation is obtained from the fit to the data in the considered categories. As such, the only relevant uncertainty originates from the covariance matrix of the dijet function fit, as indicated by the shaded region in Figs.~\ref{fig:wr} and \ref{fig:zr}.

The dominant uncertainties in the signal arise from the \PH jet and \V jet tagging. The b tagging scale factor uncertainties~\cite{CMS-PAS-BTV-15-002} are varied by one standard deviation, and the difference in the signal yield is estimated to be 4--8\% for the tight categories and 2--5\% for the loose categories. The same procedure is applied to the \nsub scale factors, whose uncertainty is measured to be 11\% for the high purity and 23\% for the low purity category, as reported in Section~\ref{sec:eventreconstruction}. The uncertainties associated with the Higgs boson mass selection and the \V jet tagging extrapolation from the \ttbar scale to larger jet \pt are estimated by using an alternative \HERWIGpp~\cite{Bahr:2008pv} shower model, and are found to be 5--7\% and 3--20\% for the \PH and \V jet candidates, respectively. Both b tagging and \nsub uncertainties are anti-correlated between the corresponding categories.

Uncertainties in the reconstruction of the hadronic jets affect both the signal efficiency and the shape of the reconstructed resonance mass. The four-momenta of the reconstructed jets are scaled and smeared according to the uncertainties in the jet \pt and momentum resolution. These effects account for a 1\% uncertainty in the mean and a 2\% uncertainty in the width of the signal Gaussian core. The jet mass is also scaled and smeared according to the measurement of the jet mass scale (resolution), giving rise to 2\% (12\%) normalization uncertainties, respectively, and up to 16\% (18\%) migration effects between the \PW and \Z mass regions depending on the category and signal hypothesis.

Additional systematic uncertainties affecting the signal normalization include the lepton identification, isolation and missing transverse momentum vetoes (accounting for 1\% each), pileup modeling (0.1\%), the integrated luminosity (2.5\%)~\cite{CMS:lumi}, and the choice of the PDF set~\cite{Butterworth:2015oua} (1\% for acceptance, 6--25\% for the normalization). The factorization and renormalization scale uncertainties are estimated by varying the scales up and down by a factor of 2, and the resulting effect is a variation of 4--13\% in the normalization of the signal events.

\section{Results and interpretation}
\label{sec:res}

Results are obtained by fitting the background functions and the signal shape to the unbinned data \mVH distributions in the eight categories. In the fit, which is based on a profile likelihood, the shape parameters and the normalization of the background in each category are free to float. Systematic uncertainties are treated as nuisance parameters and are profiled in the statistical interpretation~\cite{CMS-NOTE-2011-005}. The background-only hypothesis is tested against the signal hypothesis in the eight exclusive categories simultaneously.
The asymptotic modified frequentist method~\cite{Asymptotic} is used to determine limits at 95\% CL on the contribution from signal~\cite{CLS1,CLS2}.
Limits are derived on the product of the cross section for a heavy vector boson \X and the branching fractions for the decays $\X\to\VH$ and $\htobb$, denoted $\sigma(\X) \, \B(\X\to\VH) \, \B(\htobb)$.

\begin{figure}[!htb]\centering
    \includegraphics[width=0.495\textwidth]{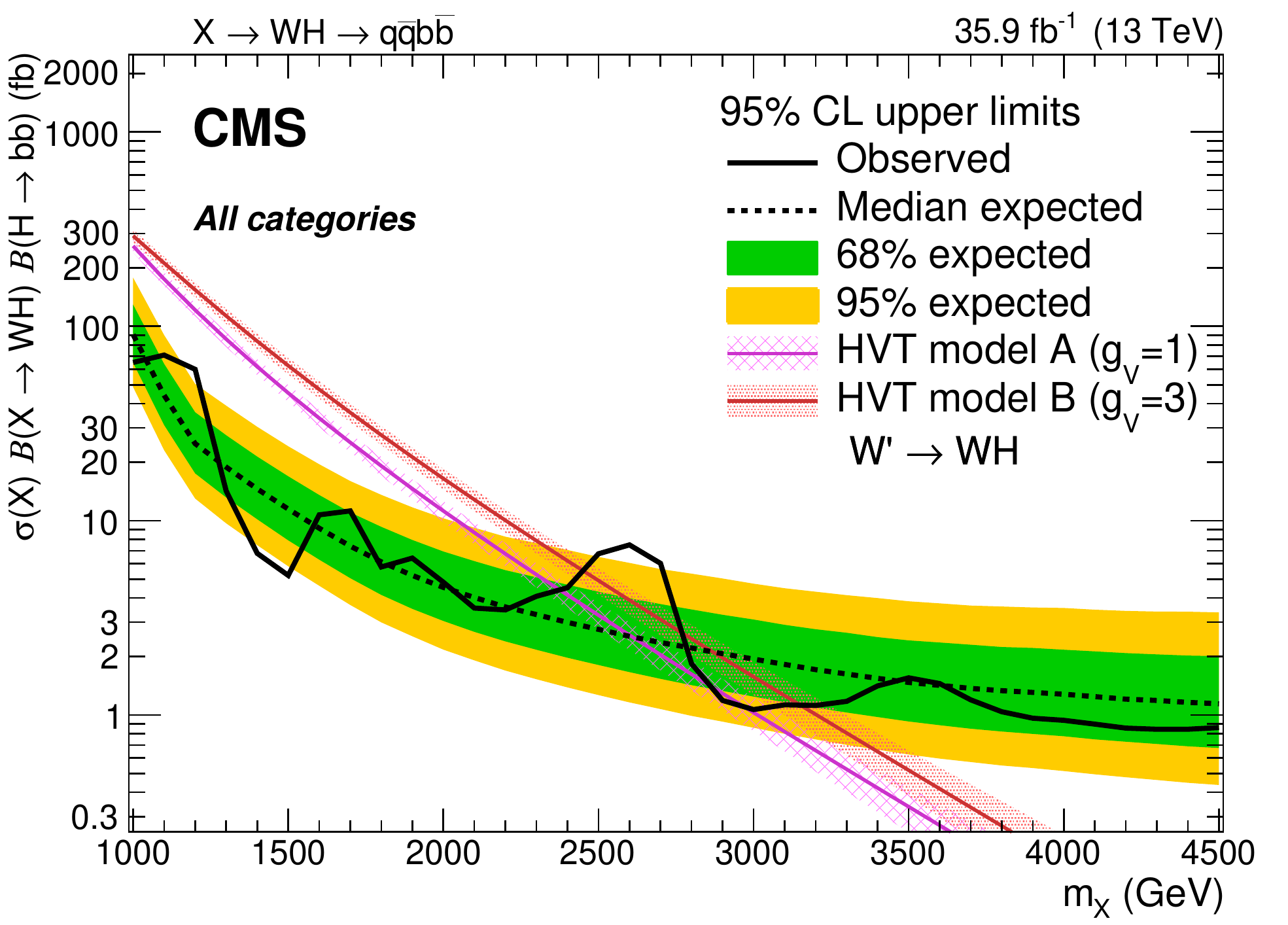}
    \includegraphics[width=0.495\textwidth]{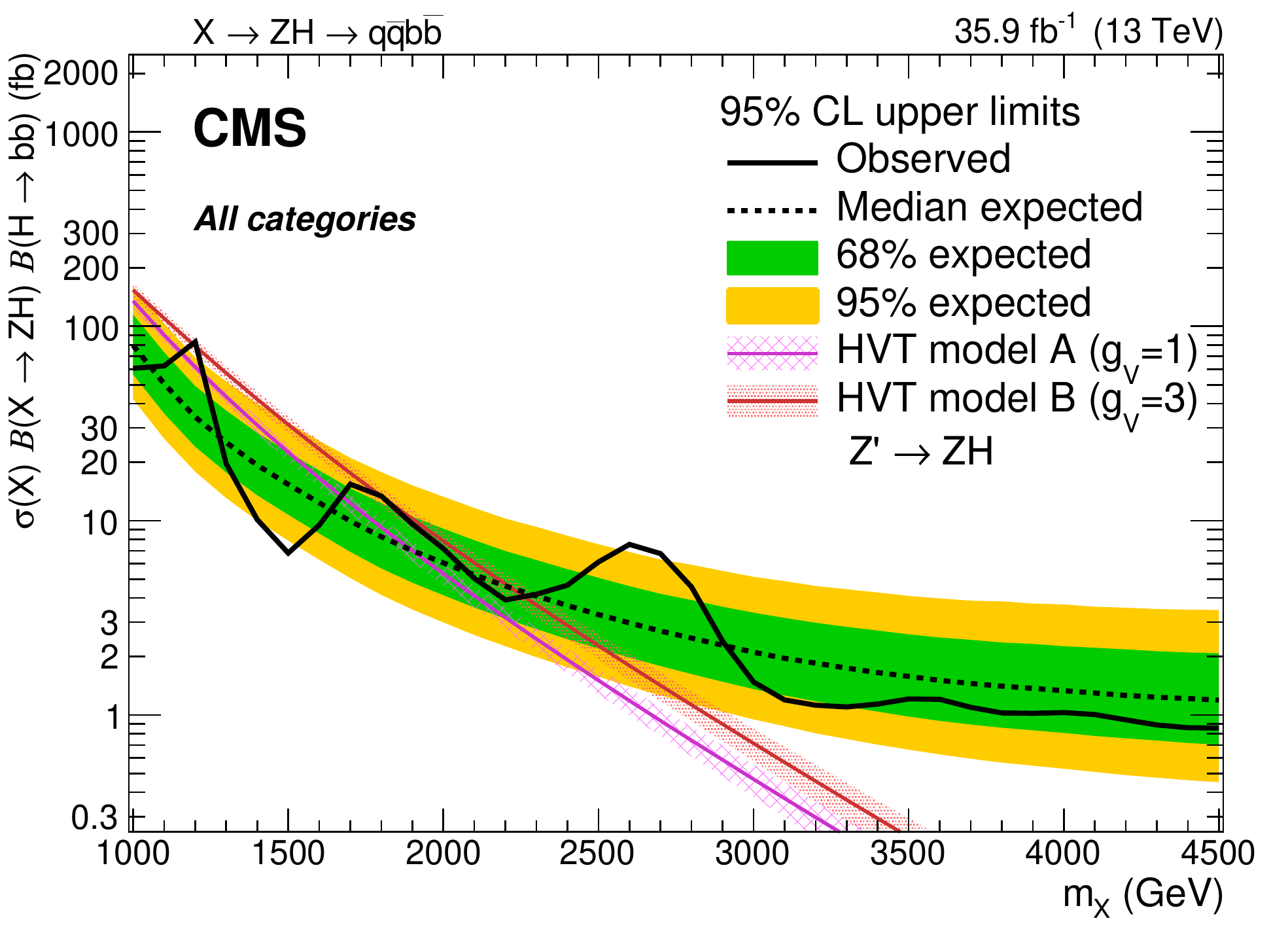}
    \caption{Observed and expected 95\% CL upper limits on the product $\sigma(\X) \, \B(\X\to\PW\PH) \, \B(\htobb)$ (\cmsLeft) and $\sigma(\X) \, \B(\X\to\Z\PH) \, \B(\htobb)$ (\cmsRight) as a function of the resonance mass for a single narrow spin-1 resonance, for the combination of the eight categories, and including all statistical and systematic uncertainties. The inner green and outer yellow bands represent the ${\pm}1$ and ${\pm}2$ standard deviation uncertainties in the expected limit. The purple and red solid curves correspond to the cross sections predicted by the HVT model~A and model~B, respectively.}
  \label{fig:limit}
\end{figure}

\begin{figure}[!htb]\centering
    \includegraphics[width=\cmsFigWidth]{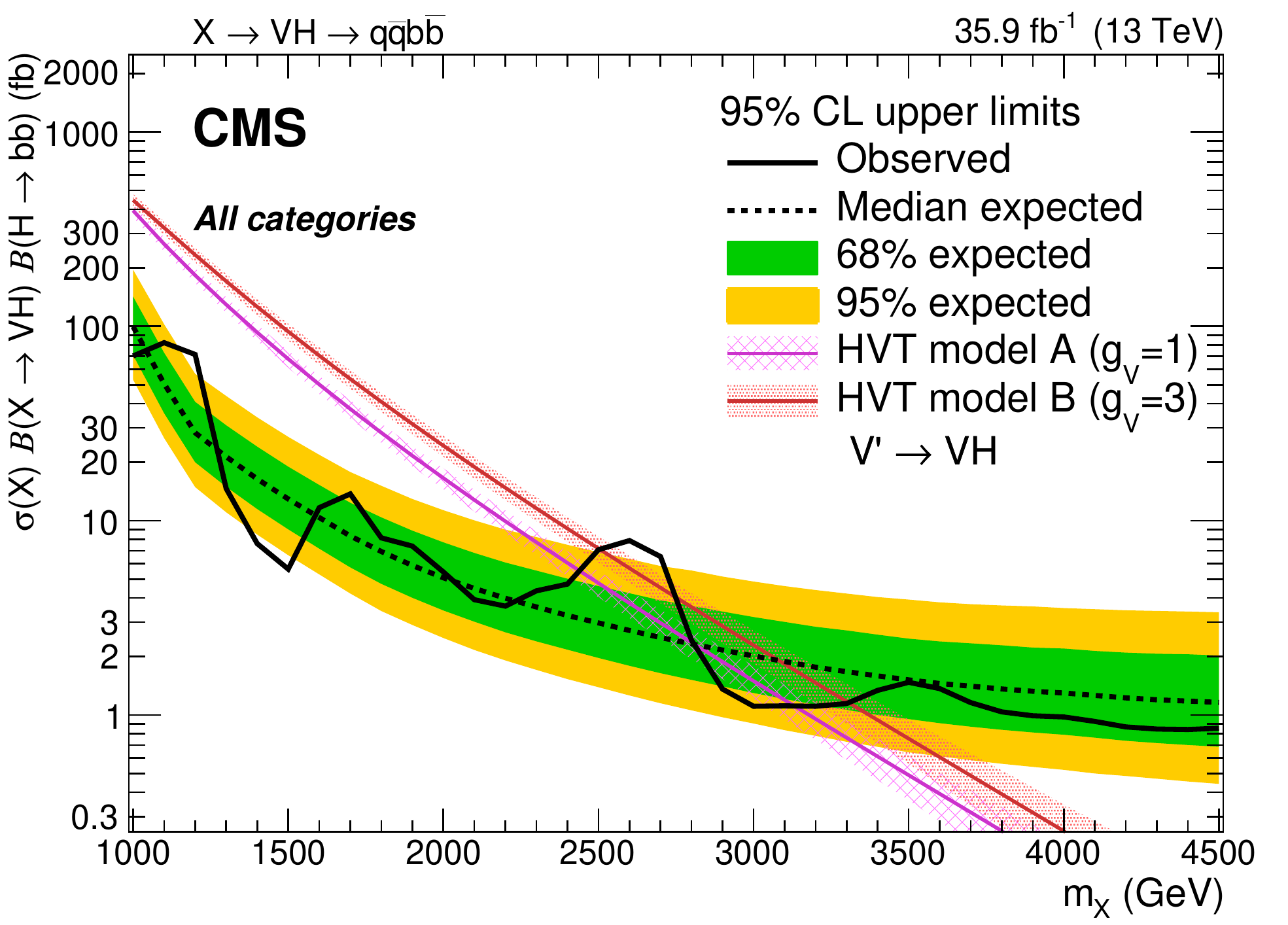}
    \caption{Observed and expected 95\% CL upper limits with the ${\pm}1$ and ${\pm}2$ standard deviation uncertainty bands on the product $\sigma(\X) \, \B(\X\to\V\PH) \, \B(\htobb)$ in the combined heavy vector triplet hypothesis, for the combination of the eight categories. The purple and red solid curves correspond to the cross sections predicted by the HVT model~A and model~B, respectively.}
  \label{fig:xvh}
\end{figure}

Results are given in the spin-1 hypothesis both for $\PWpr\to\PW\PH$ and $\PZpr\to\Z\PH$ separately (Fig.~\ref{fig:limit}) as well as for the heavy vector triplet hypothesis $\V'\to\V\PH$ summing the mass-degenerate $\PWpr$ and $\PZpr$ production cross sections together (Fig.~\ref{fig:xvh}), where they are compared to the cross sections expected in HVT models~A and~B. Upper limits in the range 0.9--90\unit{fb} are set on the product of the cross section and the combined branching fraction for its decay to a vector boson and a Higgs boson decaying into a pair of b quarks, and compared to the  HVT models~A and~B. In this case, the value of $\B(\htobb)$ is assumed to be $0.5824 \pm 0.008$~\cite{deFlorian:2016spz}. The uncertainties in the signal normalization from PDFs, and factorization and renormalization scales, are not profiled in the likelihood fit, as they are reported separately as uncertainties in the model cross section. From the combination of the eight categories, a narrow \PWpr resonance with $m_{\PWpr} < 2.37 \TeV$ and $2.87 < m_{\PWpr} < 2.97 \TeV$ can be excluded at 95\% CL in model~A, and $m_{\PWpr} < 3.15\TeV$ except in a region between 2.45 and 2.78\TeV in model~B. A \PZpr resonance with $m_{\PZpr} < 1.15\TeV$ or $1.25 < m_{\PZpr} < 1.67\TeV$ is excluded in the HVT model~A, and the ranges $m_{\PZpr} < 1.19\TeV$ and $1.21 < m_{\PZpr} < 2.26\TeV$ are excluded in model~B.

The excluded regions for the HVT masses are 1.00--2.43\TeV and 2.81--3.13\TeV in the benchmark model~A. The ranges excluded in the framework of model~B are 1.00--2.50 and 2.76--3.30\TeV, significantly extending the reach with respect to the previous $\sqrt{s}=8\TeV$ and $\sqrt{s}=13\TeV$ searches~\cite{Khachatryan:2015bma,Khachatryan2017137}. The largest observed excess, according to the modified frequentist CLs method~\cite{CLS2}, corresponds to a mass of 2.6\TeV and has a local (global) significance of 2.6 (0.9) standard deviations.

\begin{figure}[!htb]\centering
    \includegraphics[width=\cmsFigWidth]{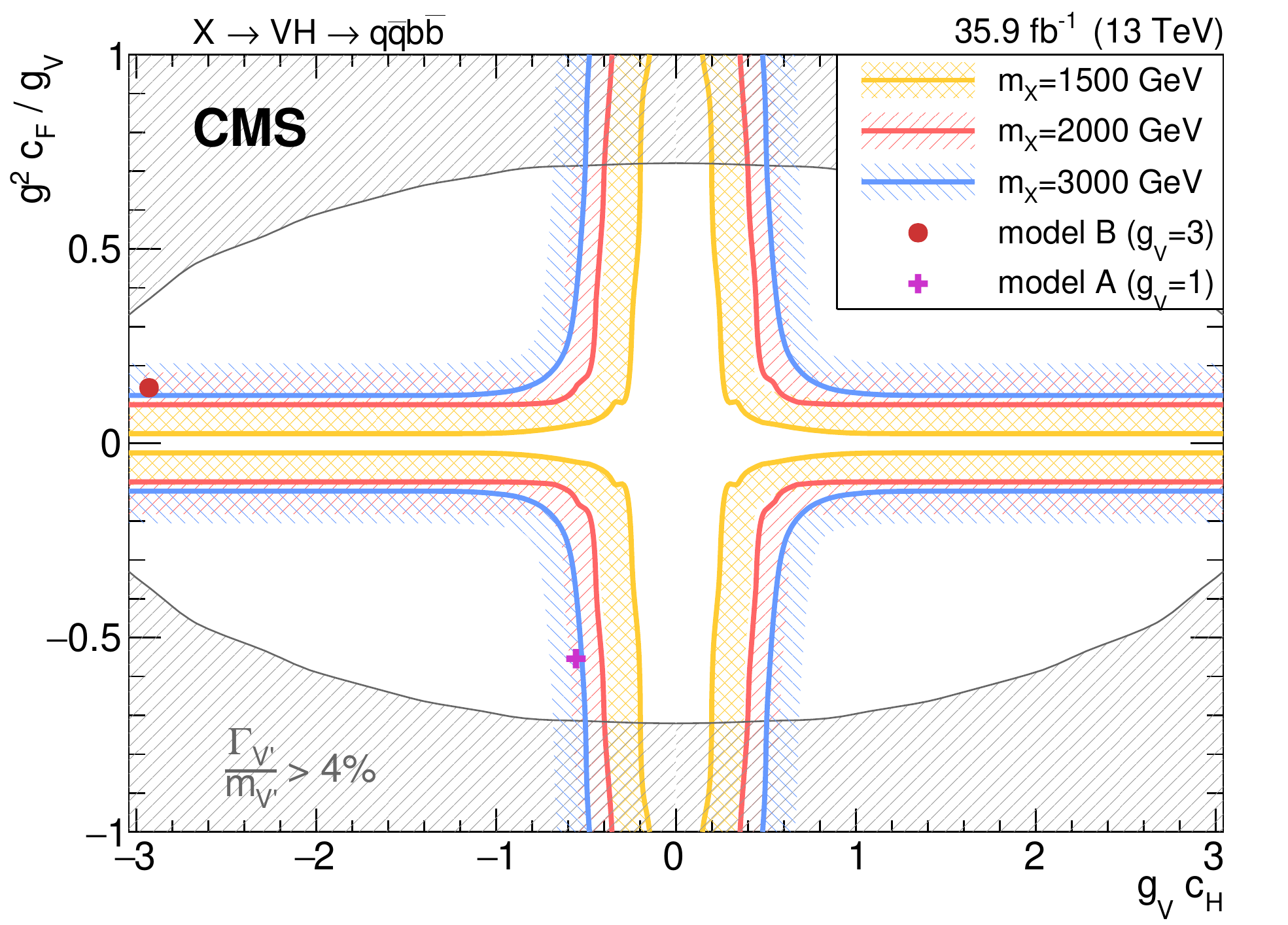}
    \caption{Observed exclusion in the HVT parameter plane $\left[ \gV \cH, \ g^2 \cF / \gV \right]$ for three different resonance masses (1.5, 2.0, and 3.0\TeV). The parameter $\gV$ represents the coupling strength of the new interaction, $\cH$ the coupling between the HVT bosons and the Higgs boson and longitudinally polarized SM vector bosons, and $\cF$ the coupling between the heavy vector bosons and the SM fermions. The benchmark scenarios corresponding to HVT model~A and model~B are represented by a purple cross and a red point. The gray shaded areas correspond to the region where the resonance natural width is predicted to be larger than the typical experimental resolution (4\%) and thus the narrow-width approximation does not apply.}
  \label{fig:hvt}
\end{figure}

The exclusion limit shown in Fig.~\ref{fig:xvh} can be interpreted as a function of the coupling strength of the heavy vectors to the SM bosons and fermions in the $\left[ \gV \cH, \ g^2 \cF / \gV \right]$ plane. Here, the uncertainties in the signal normalization from PDFs, and factorization and renormalization scales, are profiled in the fit. The excluded region of the parameter space for narrow resonances determined with an analysis of the combined eight categories of data is shown in Fig.~\ref{fig:hvt}. The region of the parameter space where the natural width of the resonances exceeds the typical experimental width of 4\%, and thus invalidates the narrow width approximation, is also indicated in Fig.~\ref{fig:hvt}.

\section{Summary}\label{sec:conclusions}

A search for a heavy resonance with a mass above 1\TeV and decaying into a vector boson and a Higgs boson, has been presented. The search is based on the final states associated
with the hadronic decay modes of the vector boson and the decay mode of the Higgs boson to a \bbbar pair. The data sample was collected by the CMS experiment at $\sqrt{s}=13\TeV$ during 2016, and corresponds to an integrated luminosity of 35.9\fbinv.
Within the framework of the heavy vector triplet model, mass-dependent upper limits in the range 0.9--90\unit{fb} are set on the product of the cross section for production of a narrow spin-1 resonance and the combined branching fraction for its decay to a vector boson and a Higgs boson decaying into a pair of b quarks.
Compared to previous measurements, the range of resonance masses excluded within the framework of benchmark model~B of the heavy vector triplet model is extended substantially to values as high as 3.3\TeV. More generally, the results lead to a significant reduction in the allowed parameter space for heavy vector triplet models.

\begin{acknowledgments}
We congratulate our colleagues in the CERN accelerator departments for the excellent performance of the LHC and thank the technical and administrative staffs at CERN and at other CMS institutes for their contributions to the success of the CMS effort. In addition, we gratefully acknowledge the computing centers and personnel of the Worldwide LHC Computing Grid for delivering so effectively the computing infrastructure essential to our analyses. Finally, we acknowledge the enduring support for the construction and operation of the LHC and the CMS detector provided by the following funding agencies: BMWFW and FWF (Austria); FNRS and FWO (Belgium); CNPq, CAPES, FAPERJ, and FAPESP (Brazil); MES (Bulgaria); CERN; CAS, MoST, and NSFC (China); COLCIENCIAS (Colombia); MSES and CSF (Croatia); RPF (Cyprus); SENESCYT (Ecuador); MoER, ERC IUT, and ERDF (Estonia); Academy of Finland, MEC, and HIP (Finland); CEA and CNRS/IN2P3 (France); BMBF, DFG, and HGF (Germany); GSRT (Greece); OTKA and NIH (Hungary); DAE and DST (India); IPM (Iran); SFI (Ireland); INFN (Italy); MSIP and NRF (Republic of Korea); LAS (Lithuania); MOE and UM (Malaysia); BUAP, CINVESTAV, CONACYT, LNS, SEP, and UASLP-FAI (Mexico); MBIE (New Zealand); PAEC (Pakistan); MSHE and NSC (Poland); FCT (Portugal); JINR (Dubna); MON, RosAtom, RAS, RFBR and RAEP (Russia); MESTD (Serbia); SEIDI, CPAN, PCTI and FEDER (Spain); Swiss Funding Agencies (Switzerland); MST (Taipei); ThEPCenter, IPST, STAR, and NSTDA (Thailand); TUBITAK and TAEK (Turkey); NASU and SFFR (Ukraine); STFC (United Kingdom); DOE and NSF (USA).

\hyphenation{Rachada-pisek} Individuals have received support from the Marie-Curie program and the European Research Council and Horizon 2020 Grant, contract No. 675440 (European Union); the Leventis Foundation; the A. P. Sloan Foundation; the Alexander von Humboldt Foundation; the Belgian Federal Science Policy Office; the Fonds pour la Formation \`a la Recherche dans l'Industrie et dans l'Agriculture (FRIA-Belgium); the Agentschap voor Innovatie door Wetenschap en Technologie (IWT-Belgium); the Ministry of Education, Youth and Sports (MEYS) of the Czech Republic; the Council of Science and Industrial Research, India; the HOMING PLUS program of the Foundation for Polish Science, cofinanced from European Union, Regional Development Fund, the Mobility Plus program of the Ministry of Science and Higher Education, the National Science Center (Poland), contracts Harmonia 2014/14/M/ST2/00428, Opus 2014/13/B/ST2/02543, 2014/15/B/ST2/03998, and 2015/19/B/ST2/02861, Sonata-bis 2012/07/E/ST2/01406; the National Priorities Research Program by Qatar National Research Fund; the Programa Clar\'in-COFUND del Principado de Asturias; the Thalis and Aristeia programs cofinanced by EU-ESF and the Greek NSRF; the Rachadapisek Sompot Fund for Postdoctoral Fellowship, Chulalongkorn University and the Chulalongkorn Academic into Its 2nd Century Project Advancement Project (Thailand); and the Welch Foundation, contract C-1845. \end{acknowledgments}

\bibliography{auto_generated}
\cleardoublepage \appendix\section{The CMS Collaboration \label{app:collab}}\begin{sloppypar}\hyphenpenalty=5000\widowpenalty=500\clubpenalty=5000\textbf{Yerevan Physics Institute,  Yerevan,  Armenia}\\*[0pt]
A.M.~Sirunyan, A.~Tumasyan
\vskip\cmsinstskip
\textbf{Institut f\"{u}r Hochenergiephysik,  Wien,  Austria}\\*[0pt]
W.~Adam, F.~Ambrogi, E.~Asilar, T.~Bergauer, J.~Brandstetter, E.~Brondolin, M.~Dragicevic, J.~Er\"{o}, M.~Flechl, M.~Friedl, R.~Fr\"{u}hwirth\cmsAuthorMark{1}, V.M.~Ghete, J.~Grossmann, J.~Hrubec, M.~Jeitler\cmsAuthorMark{1}, A.~K\"{o}nig, N.~Krammer, I.~Kr\"{a}tschmer, D.~Liko, T.~Madlener, I.~Mikulec, E.~Pree, D.~Rabady, N.~Rad, H.~Rohringer, J.~Schieck\cmsAuthorMark{1}, R.~Sch\"{o}fbeck, M.~Spanring, D.~Spitzbart, J.~Strauss, W.~Waltenberger, J.~Wittmann, C.-E.~Wulz\cmsAuthorMark{1}, M.~Zarucki
% ...
\vskip\cmsinstskip
% ...
1:~~Also at Vienna University of Technology, Vienna, Austria\\
% ...

\end{sloppypar}
\end{document}